# The role of PHF5A in cancer: A review and update

Patrick Diaba-Nuhoho [1,*]
[1] Department of Paediatric and Adolescent Medicine, Paediatric Haematology and Oncology, University Hospital Münster, Germany
* Correspondence: diabanuh@uni-muenster.de

**Abstract:** PHF5A is a member of the zinc-finger proteins. To advance knowledge on their role in carcinogenesis, data from experimental studies, animal models and clinical studies in different tumorigenesis have been reviewed. Furthermore, PHF5A as an oncogenic function, is frequently expressed in tumor cells and a potential prognostic marker for different cancers. PHF5A is implicated in the regulation of cancer cell proliferation, invasion, migration and metastasis. Knockdown of PHF5A prevented the invasion and metastasis of tumor cells. Here, the role of PHF5A in different cancers and their possible mechanism in relation to recent literature is reviewed and discussed. However, there is an open promising perspective to their therapeutic management for different cancer types.

## 1. Introduction

Cancer cells take advantage of alternative splicing (AS), a posttranslational gene regulation generating RNA transcripts to produce proteins that promote tumor growth and survival [1, 2]. During AS, plant homeodomain (PHD)-finger protein 5A (PHF5A) encodes a subunit of splicing factor 3b (SF3b) complex which is important for SF3b spliceosome stability and linkage of the spliceosome to histones [3, 4]. PHF5A is highly conserved, expressed in the nucleus, regulating pluripotency, cell differentiation of embryonic stem cells through transcription processes [5, 6]. In addition, it regulates glioma pleomorphic stem cell viability via the cell cycle [7]. Also, PHF5A regulates DNA repairs in class switch recombination via p400 and histone H2A variant deposition [8]. Experimental data suggest PHF5A can play oncogenic role and has been reported to promote tumor progression in several cancer types [9, 4, 10-14]. Taken together, insight into the potential mechanism of PHD inhibitors and in particular, PHF5A action may be a promising therapeutic potential in cancer. Here the importance of PHF5A in cancer is described, its potential role in different cancers and possible effect of PHF5A as a therapeutic target is reviewed.

## 2. PHF5A as an oncoprotein

PHF5A belongs to a superfamily of PHD-finger genes which is a highly conserved small transcriptional regulator involved in pre-mRNA splicing, encoding a protein of 110 amino acid with the PHD zinc finger domain [15]. It has a protein molecular mass of 14kDa [16] and was first described by Schindler and colleagues from *Arabidopsis thaliana* as HAT3.1, a homeodomain protein containing conserved cysteine-rich region which are known to be present in yeast, plants and mammals [17]. By binding to the promoter of the Connexin 43 gene in rats, PHF5A acts as a specific coactivator that upregulates the expression of Connexin 43 gene in response to estrogen induction [18]. PHF5A can be found in two major groups of proteins, one as transcriptional activators, repressors or cofactors and the other as chromatin-associated proteins or chromatin modulating complexes such as acetyltransferases or complexes containing acetyltransferases [15], and could enhance interaction with specific histone marks on chromatin-bound nucleosomes in other PHD family [19, 20]. PHF5A is an important component of the SF3b complex, directly involved in protein-protein interaction and downstream regulation of genes [16, 21, 3]. For instance, PHF5A is known to form a bridge between splicing proteins such that the N-terminal domain of PHF5A is associated with E1A-binding protein 400 (EP400) and DEAD-box helicase 1 (DDX1) whereas the C-terminal region associates with U2 small nuclear RNA auxiliary factor 1 (U2AF1) and splicing factor, arginine/serine-rich 5 (SFRS5) [21]. Furthermore, a splicing modulator that targets SF3B1 in the SF3b complex could inhibit the global action of splicing modulators in both WT- and Y36C PHF5A expressing cells acting at the branch point adenosine binding pocket by forming a PHF5A-SF3b complex [3]. Suggesting that there is a common binding site at the branch point adenosine pocket and PHF5A is a key node of interaction at this branch point for small molecule splicing modulators [3]. In a recent study that analysed human tissue specific expression by RNA-sequencing and antibody-based profiling, tissue samples of 95 human individuals representing 27 different tissues were sequenced to determine tissue specificity of protein-coding genes [22]. Tissue-specificity expression levels of PHF5A could be detected in several organs after functional analysis, with abundant levels in the bone marrow and the lymph nodes [22]. In lung adenocarcinoma (LAC) patients, PHF5A expression correlated with poor survival, tumor size and lymph node metastasis [12]. Importantly, it is essential for the maintenance of pluripotency and cellular reprogramming by directing the transcriptional program and further required for normal exon recognition in glioblastoma stem cells to maintain cell expansion and viability [6, 7]. Overexpression of PHF5A is associated with proliferation and migration in normal embryonic stem cells [6] and tumor cells such as lung cancer [12, 13], non-small cell lung cancer [23, 24], colorectal cancer [14], breast cancer [4], brain tumor [7] and pancreatic tumor [11]. Recent studies have identified PHF5A as a predictive marker for poor prognosis in oral squamous cell carcinoma [25], non-small cell lung cancer [12], colorectal cancer [26, 14], breast cancer [4], liver cancer [10] and gastric cancer [9]. Thus, in both basic and essential cellular functions, such as cancer development, PHF5A protein is important. Taken together, insight into the potential mechanism of PHD inhibitors and in particular, PHF5A action may be a promising therapeutic potential in cancer.

## 3. PHF5A in different pathways

PHF5A, a protein involved in alternative splicing and various molecular interactions, plays a multifaceted role in cancer progression. Its impact on alternative splicing processes affects the diversity of protein variants, potentially contributing to tumor progression, aging, apoptosis, and cell proliferation. Furthermore, PHF5A interacts with signaling pathways such as IGF-1, RhoA/ROCK, NF-κB, and PAF1, influencing gene expression, cell cycle regulation, cytoskeleton maintenance, inflammation, and transcriptional elongation.

### *3.1. PHF5A in alternative splicing and histone deacetylases*

Alternative splicing (AS) is a post transcriptional regulator of genes where a single gene generates multiple RNA transcripts leading to proteomic diversity [27]. Abnormal splicing is associated with pathological and oncological processes such as tumor progression [28, 13], aging [29], apoptosis [30], proliferation and migration [26]. PHF5A as a component of the splicing factor complex SF3b acting at the branch point adenosine binding pocket is important for recognising exons [7, 3]. By participating in spliceosome formation together with splicing factors 3b, 3a, and 12S RNA, PHF5A maintains the stability of SF3b spliceosome [14]. In brain tumor initiating cells, PHF5A is required for proliferation and expansion by maintaining exon recognition and its knockdown in glioblastoma multiforme stem cells could inhibit splicing [7]. More importantly during mouse spermatogenesis, it forms a bridge protein between splicing proteins and ATP-dependent helicases enabling interaction with splicing factors U2AF1 and SFRS5 and helicases EP400 and DDX1 [21]. Indicating the essential role of PHF5A in cellular function. However, in certain cancers such as breast cancer, colorectal cancer, lung cancer, PHF5A promotes tumor progression and regulates the splicing of multiple genes [4, 14, 26, 13]. For instance, in fibroblast cell lines, PHF5A decrotonylation via AS could accelerate fibroblast senescence in aging progress. Sirtuin 7 (SIRT7), an $NAD^+$-dependent protein deacetylase induced PHF5A decrotonylation at k25 via AS with intron inclusion to downregulate cyclin-dependent kinase 2 (CDK2) expression, a cell-cycle regulator gene [29]. Further dysregulation of AS by PHF5A in colorectal cancer (CRC) induced TEA domain transcription factor 2 (TEAD2) exon 2 inclusion to activate yes-associated protein (YAP) signaling promoting cancer progression. Thus, abnormal splicing of TEAD2 is responsible for PHF5A promoted proliferation and migration in CRC [26]. Since acetylation of PHF5A enhances U2 small nuclear ribonucleoprotein (snRNP) interaction [14], PHF5A hyperacetylation with reduced intron inclusion via alternative splicing could stabilize lysine demethylase 3A (KDM3A) mRNA to increase its protein expression in CRC. PHF5A can also promote progression of lung cancer through AS leading to gene dysregulation involved in cell cycle (s-phase kinase associated protein 2 (SKP2), checkpoint kinase 2 (CHEK2), ataxia telangiectasia and Rad3-related protein (ATR) and apoptosis (AP15, B-cell lymphoma 2 like 13 (BCL2L13) [13]. In prostate cancer, PHF5A is required for efficient exon 7 (ex7) inclusion during AS, which is important for muscleblind-like 1 (MBNL1) protein homodimerization. Therefore, AS of isoforms lacking ex7 induced DNA damage and inhibited cell viability and migration. Similarly, histone deacetylases (HDACs) regulate epigenetic and non-epigenetic modifications and impact compaction of chromatin, regulation of gene transcriptional expression, proliferation, differentiation, angiogenesis, immune escapes, cell cycle arrest and apoptosis [31, 32]. The development and pluripotency of many cancer stem-like cells that are responsible for invasiveness, drug resistance, and relapse of cancers are thought to be regulated by this pattern [33]. Thus, disruption of the acetylation signature in HDACs play important roles in development and progression of cancers [34]. While HDACs are important in chromatin remodeling and as a key regulator of epigenetic modification, PHF5A could interact with DNA, RNA, proteins and bind chromatin through its PHD domain [15, 4]. Hence PHF5A may interact and regulate HDAC activity. Interestingly, PHF5A was recently found to maintain the capability of cancer stem-like cells for self-renewal and differentiation in non-small cell lung cancer via HDAC8 regulation [23]. Experimental evidence reveal HDAC8 is located downstream of PHF5A and showed a constantly decreasing trend when PHF5A was knockdown [23]. In glioblastoma multiforme stem cells, PHF5A knockdown resulted in the constitutive exon skipping of the cancer-associated deacetylase HDAC6 [7]. Taken together, these findings drive home the importance of AS in epigenetic regulation of different cancers, the pleiotropic effects of PHF5A as a regulator in the AS of essential genes in different cancers, a positive association between HDAC and PHF5A that may promote cancer progression and tumor formation and a promising anticancer therapeutic target.

### *3.2. PHF5A and IGF-1*

Insulin-like growth factor 1 (IGF1) mediates cell process, proliferation, survival and metabolism [35]. IGFs regulate the expansion of other genes implicated in angiogenesis, metastasis, vascular endothelial growth factor and E-cadherin [36]. IGFs are activated by receptor tyrosine kinases upon binding to the IGF-1 receptor initiating downstream phosphoinositide-3-kinase/protein kinase B, mitogen-activated protein kinase/epidermal growth factor receptor extracellular-regulated kinase (PI3K/AKT, MAP/ERK) signaling and play important role in tumor progression [35, 37]. In LAC, pathway enrichment and western blot confirmation indicated that PHF5A regulated the cell cycle to participate in tumor development. This at the transcription levels of IGF-binding protein 3 (IGFBP3), DNA-damage inducible transcript 3 (DDIT3) and tumor protein P53 genes were upregulated while PI3K, AKT and Skp2 were downregulated after PHF5A silencing in H1299 cell lines [12]. An indication that shows an association between PHF5A and IGF-1. Whereas, IGFBP3 is one of the six IGF-binding proteins regulating cell activity, cell proliferation, apoptosis, cell cycle, intracellular metabolism through IGF-1/IGF-1R dependent and independent pathways [38-40], both P53 and IGFBP3 are also tumor suppressors and influenced by IGF-1 [41, 42]. Also, DDIT3 is a transcription factor activated by endoplasmic reticulum (ER), unfolded proteins, amino acid deprivation and oxidative stress [43, 44]. IGF-1 can have positive and negative influence on ER enhancing their ability to stress adaptation and returning to normal homeostasis or aggravate stress and promotes apoptosis [45]. Yang et al. found that PHF5A silencing resulted in Skp2, PI3K/AKT2 and P53 downregulation. An indication suggesting the inactivation of the PI3K/AKT2 pathway resulted in down-regulating Skp2, decreasing the degradation of P53, a protein target for Skp2 ubiquitination and reducing cell proliferation and tumor progression in LAC [12]. AKT/mammalian target of rapamycin (AKT/mTOR) is important in signal transduction, apoptosis, proliferation, cell cycle and growth of tumor cells [46]. Interestingly, decreased phosphorylation of AKT and mTOR was seen after PHF5A knockdown in gastric cancer (GC) cell lines [9]. Furthermore, in GC cell lines, downregulation of PHF5A accelerates protein degradation of FOS [9], a member of the Fos family of transcription factors regulating cell proliferation and differentiation in tumor [47]. Notwithstanding, Skp2 overexpression could reduce the protein stability of FOS in GC cells, indicating that protein stability of FOS was affected by PHF5A and Skp2 [9]. However, proteasome inhibitor treatment could abolish the effect of PHF5A knockdown and Skp2 overexpression on the FOS protein stability. Showing that knockdown of PHF5A in GC cells diminished PHF5A overexpression with decreased protein stability of FOS via Skp2-mediated ubiquitination [9]. Therefore, IGF signaling is associated with different cancer types and can influence the activity of PHF5A in cancer progression.

### *3.3. PHF5A and RhoA/ROCK*

The Ras homolog family A/Rho-associated kinase (RhoA/ROCK) is a small GTPase protein with multifunctional kinases important in cell cycle control and cytoskeleton maintenance [48, 49]. They are also involved in cancer cell apoptosis, proliferation and migration [50]. Down regulation of PHF5A could reduce RhoA, ROCK and chromodomain helicase DNA-binding protein 4 (CHD4) expression in cell lung cancer (NSCLC) cell lines. CHD4, an ATP-dependent chromatin remodelling protein is involved in cell progression, differentiation and cancer development [51, 24]. PHF5A interacts with specific histone marks on chromatin-bound nucleosomes and hence can interact with CHD4 and activate RhoA/ROCK signaling pathway [24, 5]. Interestingly, CHD4 overexpression promoted proliferation and migration of non-small cell lung cancer cells in H292, and PC-9 cell lines while this effect was reduced in A549 and H1299 cell lines when CHD4 was knockout [24]. This over expression effect of CHD4 in cancer was attributed to its interaction with PHF5A. Indeed, further experimental evidence by Xu et al. targeting PHF5A using small interfering RNAs (siRNAs) in A549 cell lines revealed a down regulation of PHF5A, reduced RhoA, ROCK and CHD4. Again, a co-IP in A549 whole cell lysates using antibodies against CHD4 and PHF5A identified an interaction suggesting a potential of PHF5A binding to CHD4 [24]. This indicates the possibility of PHF5A regulating downstream targets of RhoA/ROCK signaling pathway via association with certain chromatin remodelling proteins.

### *3.4. PHF5A and NF-κB*

Nuclear factor-κB (NF-κB) signaling is a transcription factor that plays an important role in inflammation, proliferation, survival, invasion, angiogenesis and metastasis of cancer cells [52, 53]. Interestingly, Yang et al. [10] showed that PHF5A plays an important role in the activation of NF-κB signaling during their overexpression in hepatocellular carcinoma (HCC) cells and tissue. Furthermore, inhibition of PHF5A or NF-κB signaling could significantly reduce HCC progression. Importantly, knockdown of PHF5A downregulated downstream targets of NF-κB signaling such as p65, inhibitor of NF-κB (IκBα) and IκB kinase (IKK-β) phosphorylation and matrix metallopeptidase 9 (MMP9) and snail family transcriptional repressor 2 (SLUG) proteins [10]. Therefore, PHF5A association with NF-κB signaling as a result of inflammation can promote cancer progression.

### 3.5. PHF5A and Paf1

PAF1, an RNA polymerase II-associated factor 1 regulates transcription, initiation and elongation and is a component of the PAF1 complex (PAF1C) [54]. PHF5A is important for PAF1 complex recruitment, stabilization, release and elongation [6, 55]. The interaction of PAF1 with PHF5A activates pluripotent genes in the embryonic stem cells to control the elongation of RNA polymerase II [6]. Importantly in embryonic stem cells, loss of PHF5A enhanced RNA polymerase pausing and decreased Serine-2-phosphorylation of polymerase II (Ser-2-P-Pol II) levels in the promoter regions of PAF1 targets to reduce downstream gene expression [6]. Recent interactions of PHF5A and PAF1 has been demonstrated in pancreatic cancers [56, 11]. Furthermore, in pancreatic cancer stem cells, PAF1 interacts with DEAD-box helicase 3 (DDX3) and PHF5A at the promoter region of Nanog to regulate NANOG expression [11], a master regulator of pluripotency [57]. It would be interesting to decipher how PAF interaction with PHF5A exacerbate cancer growth via their synergistic action in different cancer models.

Therefore, these different pathways (Figure 1) by which PHF5A promotes cancer progression and regulates multiple genes may present a promising target for cancer therapy.

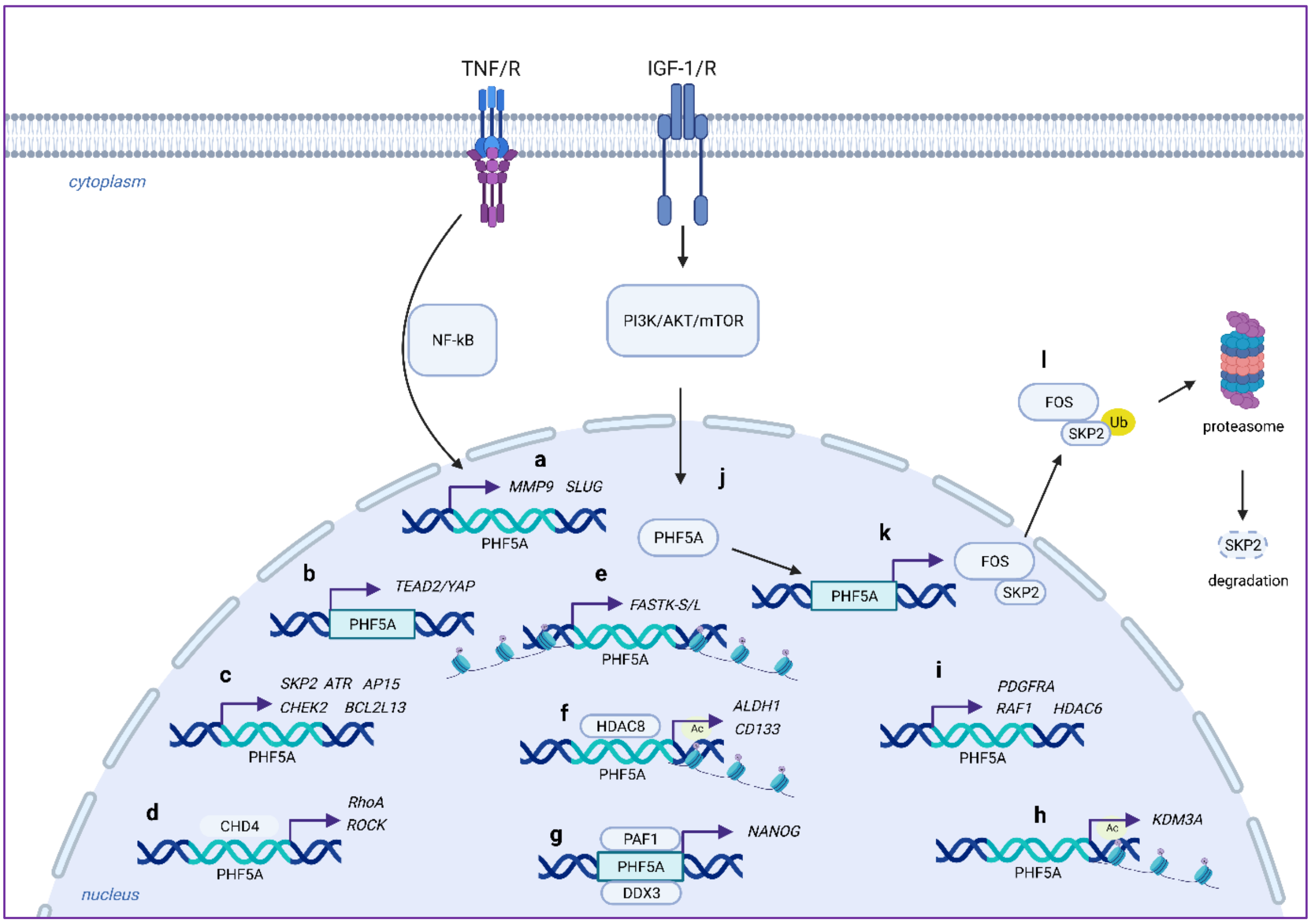

**Figure 1.** Schematic diagram of PHF5A involvement in tumor growth

(a) PHF5A regulates expression of genes *MMP9* and *SLUG* via NF-kB signaling. (b) PHF5A promotes cancer progression via alternative splicing (AS) of *TEAD2* to active *YAP* signaling. (c) PHF5A expression promotes cancer cell proliferation by AS of cell cycle (*SKP2*, *CHEK2*, *ATR*) and apoptotic genes (*AP15* and *BCL2L13*). (d) Both CHD4 and PHF5A interact and regulates expression of genes in the *RhoA* and *ROCK*. (e) PHF5A associates with histones and represses apoptosis via AS switching to a long truncated *FASTK-L*. (f) PHF5A increases *CD133* and *ALDH1* expressions by recruiting HDAC8 to the promoter to maintain stemness of cancer cells. (g) PHF5A interacts with PAF1 and DDX to increase *NANOG* expression and promote tumor progression. (h) Acetylation of PHF5A via AS upregulates *KDM3A* expression to promotes cancer growth. (i) AS of PHF5A with HDAC6 exon inclusion leads to gene expression of *PDGFRA*, *RAF1* and *HDAC6* to increase cancer migration. (J,k,l) IGF-1 induces the expression of PHF5A to increase FOS expression, FOS interacts with SKP2 to induce proteasomal degradation and impair SKP2 to exert its function. Created with BioRender.com.

## 4. PHF5A in specific cancers

Most human cancers show an upregulation of PHF5A. Aberrant expression in both *in-vivo* and *in vitro* studies has been indicated, with many epigenetic and non-epigenetic regulations impacted by aberrant splicing and other factors [9, 4, 10-14]. Over expression of PHF5A is strongly indicated in several cancer types [9, 4, 10-14]. Similarly, other studies have shown the importance of PHF5A in regulating cell cycle, maintaining stem cell pluripotency and differentiation (6, 7). PHF5A-regulated genes in the IGF-1 pathway led to tumor survival, proliferation and growth in lung cancer [12]. Recently, PHF5A was found to maintain cancer progression in lungs via multiple signaling pathways involving histone deacetylase 8 (HDAC8) and IGF-1 upregulation [23, 12]. Following PHF5A silencing IGFBP3, DDIT3 and tumor protein P53 were markedly upregulated at the transcription level, while PI3K, AKT2 and Skp2 were downregulated [12]. These are important classes of mitogens involved in the IGF-1/IGF-1R pathway (in)directly and could induce or suppress cell growth and proliferation [58-60, 12]. By binding to the receptors on insulin, IGFs such as IGF-1 receptors, receptor tyrosine kinases are activated and initiate the downstream PI3K/AKT signaling pathway to induce tumor progression [61, 36]. Therefore, several oncogenic signaling involving AS and acetylation, IGF-PI3/AKT/mTOR pathways, RhoA/ROCK signaling, NF-kB and the PAF1 complex are reported to be affected during PHF5A knockdown [10, 13, 24, 11, 9]. In Table 1, the findings of specific cancer types in experimental, animal and clinal studies is reported. An indication of the various cancer types and their association with PHF5A is indicated while their therapeutic targets are highlighted in Table 2.

### *4.1. PHF5A in endometrial cancer*

PHF5A is upregulated in the ovary tissue of the murine model acting as a transcription factor in the expression of the Connexin 43 gene [15]. Connexins are membrane proteins forming channels involved in cell-to-cell communication [62-64]. They are thought to be upregulated by estrogen abundant in the uterus (myometrium) and important for coordination of contraction during end of gestation [18]. PHF5A binding to the promoter region of the Connexin 43 gene promotes its up regulation by estrogen induction [18]. However, endometrial cancer development has direct effect on the levels of PHF5A expressions. For instance, PHF5A expression in endometrial adenocarcinoma was increased in murine and human studies when compared with that of benign samples upon assessing the association of PHF5A with tumors [65]. Indicating that aberrations in tumors lead to changes in gene expression and therefore may be involved in endometrial cancer development. Yet the mechanism of how PHF5A contributes to endometrial cancer still needs further investigation.

### *4.2. PHF5A in brain cancer*

PHF5A is expressed in brain tumour specimen, glioma cells and is localised in the nucleus [7]. To identify genes differentially required for patient-derived glioblastoma multiforme stem cells (GSCs) expansion, Hubert and colleagues screened GSC patient isolates and neural stem cells (NSCs) using genome-wide RNAi. They showed a Knockdown of PHF5A disrupted splicing of multiple essential genes, induced cell cycle arrest and loss of viability in glioblastoma stem cells [7]. Further indicating the requirement of PHF5A for initiation, maintenance and expansion of glioblastoma stem cells (GBC) and that PHF5A may be a general transcription regulator for different genes in tumor development [7]. As a proof-of principle, cell-based assays *in vitro* and during cell growth in patient-derived tumors could show that PHF5A/U2 snRNP perturbation led to G2/M arrest and growth inhibition in GSCs [7]. It has been known that PHF5A is a splicing factor and transcriptional regulator that can facilitate exon recognition [18, 16, 21]. Interestingly, Hubert et al. found a crucial role of PHF5A as a requirement for recognition of exons with distinctive C-rich 3′-splice sites in GSCs, an indication which is of critical importance for growth and maintenance of patient-derived tumors [7]. Thus, in glioblastoma, PHF5A knockdown could result in exon skipping and intron retention in hundreds of genes leading to splicing defects. For instance, Hubert et al. showed that RNA processing defects were observed in cell division cycle protein 20 (CDC20), regulator of chromosome condensation 2 (RCC2) and exons of the RTK/Ras signaling effector proto-oncogene serine/threonine-protein kinase (RAF1) and the cancer-associated deacetylase histone deacetylase 6 (HDAC6) were skipped [7]. In conclusion, the work by Hubert and colleagues, drives home the importance of PHF5A as an important therapeutic target in brain cancer management wherein splicing may be a source of tumor vulnerability.

### *4.3. PHF5A in lung cancer*

PHF5A as an oncoprotein is localised in the nucleus, regulates multiple signaling pathways associated with LAC progression and poor prognosis [12, 13]. PHF5A promoted the cell cycle transition from G0/G1 to S phase during its overexpression in H1299 and A549 cell lines [13]. PHF5A expression in LAC tissue were shown to be significantly higher than adjacent non tumor tissues (normal lung tissues) and LAC patients with high PHF5A expression showed poorer survival, positively correlated with tumor size, lymph node metastasis and clinical stage of the disease [12]. This was confirmed *in vitro* experimental assay where a knockdown of PHF5A in LAC cell line inhibited growth with decreased cell proliferation, impaired cell cycle, apoptosis and induced genome wide AS events [13, 12]. Furthermore, nude mice subcutaneously injected with H1299-Luc cells and transfected with empty or shPHF5A vector showed complete tumor formation with tumor growth, size, and weight significantly reduced in the PHF5A knockdown animals [12]. Recently, in cancer stem-like cells of non-small cell lung cancer (NSCLC), targeted knockdown of PHF5A resulted in decreased stemness phenotype and HDAC8 expression, whereas inhibition of HDAC activity affected stemness maintenance [23]. Therefore, these findings implicate PHF5A as an oncoprotein in promoting carcinogenesis and progression via AS and other signaling pathways during lung cancer.

### *4.4. PHF5A in colorectal cancer*

PHF5A is upregulated in colorectal cancers [26, 14]. In human CRC tissues, Cheng et al. demonstrate that PHF5A expression is increased, promotes proliferation, metastasis and correlates with poor prognosis [26]. Again, PHF5A was upregulated in 66% of CRC patients after immunohistochemical analysis of tissue samples, with CRC patients showing poor survival, lymph node metastasis and advanced stage of disease. Similarly, they show both *in vitro* and *in vivo* experimental models promoted CRC growth and metastasis after PHF5A overexpression. Interestingly the overexpression of PHF5A was found to be controlled by epigenetic modification via activation by histone H3-lysine-27 (H3K27) acetylation, a potent regulator of transcription located at the promoter region of PHF5A [26]. However, knockdown of PHF5A in CRC cell lines decreased proliferation and migration of CRC cells, altering splicing and affecting genes involved in RNA processing and RNA splicing. In particular knockdown of PHF5A induced alternative splicing resulting in TEAD2 exon 2 inclusion to activate YAP signaling [26]. YAP pathways are known to promote tumor growth and metastasis through transcription factors mediated by TEAD [66, 67]. Thus, activation of this pathway by TEAD2 could promote the progression of CRC. Significantly, Cheng and colleagues demonstrated in their novel finding that alternative splicing promotes cell proliferation and metastasis in CRC and evidently postulate that binding of PHF5A to TEAD2 pre-mRNA generates RNA transcripts with TEAD2 exon 2 inclusion to activate YAP signaling leading to cancer progression in CRC [26]. These findings implicate PHF5A as a strong predictive marker for tumor aggressiveness in CRC.

### *4.5. PHF5A in pancreatic cancer*

In pancreatic cancers, PHF5A is localised in the nucleus and overexpressed in patient pancreatic tissues and cell lines of cancer stem cell origin [11]. A knockdown of PAF1 and PHF5A led to a decreased expression of NANOG, indicating the Nanog promoter as a common target for both PAF1 and PHF5A. Considering the interaction of PHF5A with PAF1, more independent studies regarding the mechanism of PHF5A in pancreatic cancer is warranted.

### *4.6. PHF5A in breast cancer*

Zheng and colleague conducted a study to determine epigenetic mechanisms of PHF5A underlying breast cancer progression [4]. They determined PHF5A to be linked to poor clinical outcomes and as a prognostic predictive marker in breast cancer patients, was responsible for breast cancer proliferation, migration and tumor formation. Knockdown of PHF5A by CRISPR-based methods consistently impaired cell proliferation and migration *in vitro* and significantly reduced tumor size *in vivo* [4]. An indication that affirms the role of PHF5A as an oncoprotein in breast cancers. PHF5A was found to associate with histone H3, methylated on lysine 4, localised to promoters and regulate splicing in the genome. Knockdown of PHF5a in cell lines further induced genome wide AS events with representations of skipped exons and retained introns being frequently regulated. Thus, AS events by PHF5A induced FAS-activated serine/threonine kinase (FASTK) intron removal and suppressed apoptotic signaling pathways increasing tumor progression in breast cancer [4]. Since it was determined that PHF5A stabilizes the SF3b spliceosome complex (SF3B1-3) and binds U2AF1, a component of the splicing complex [7], knockdown of PHF5A showed decrease expression patterns in these proteins without affecting their mRNA expressions [4]. Affirming the notion of their associations with PHF5A and regulation at the protein level even in breast cancer cells. The findings from Zheng et al. are interesting, comprehensive and indicate the possibility of PHF5A target for the management of tumor in breast cancer.

### *4.7. PHF5A in liver cancer*

Yang et al. demonstrated that PHF5A is overexpressed in HCC cells and tissues, and significantly correlates with poor survival of HCC patients [10]. Suggesting that PHF5A could be a prognostic factor in HCC. Importantly HCC progression is influenced by PHF5A and its silencing in HCC cells inhibits the potential for migration and invasion of the cells [10]. Yang and colleagues found that NF-κB signaling, an important mediator of inflammation and cancer [52], is decreased in PHF5A silenced HCC cells. Evidently knockdown of PHF5A decreased the levels of p65, inhibited phosphorylation of IκB$\alpha$ and IKK-β and downregulated downstream targets of NF-κB signaling such as MMP9 and Slug. Again, blocking of NF-κB signaling in PHF5A overexpressing cells significantly weaken the stimulating effect of PHF5A on migration and invasion of HCC cells. Finally, they showed that levels of PHF5A in total cells correlated with p65 expressions in nucleus during western blot analysis of clinical HCC samples. An indication that shows the level of PHF5A associates with the activation of NF-κB signaling [10].

### *4.8. PHF5A in gastric cancer*

PHF5A has recently been shown to promote cancer progression, abundantly expressed in gastric cancer (GC) and as a poor prognostic indicator [9]. GC patient tissues and cell lines showed abundant expression of PHF5A compared to normal tissues and cells. This positively correlated with the clinical characteristics of GC such as the stage of disease, pathologic level and overall survival of patients such that as the expression levels of PHF5A increased, patient survival decreased [9]. Zheng et al. show that knockdown of PHF5A using specific siRNA (shPHF5A) in GC cell lines inhibits malignancy via AKT/mTOR signaling pathway with reduced cell proliferation, increased apoptosis, inhibition of migration in GC cells and downregulating FOS gene expression [9]. Furthermore, they indicate that knockdown of PHF5A promotes ubiquitination of FOS through the E3 ubiquitin ligase Skp2. Again, FOS knockdown in GC cells inhibited proliferation, colony formation, sphere formation and promoted apoptosis similar to the expression patterns observed in PHF5A knockout cells. A combination of FOS low expression and PHF5A overexpression in GC cells could diminish PHF5A overexpression on GC cells, implying that downregulation of FOS attenuates the promotion of PHF5A overexpression on GC cells. However, treatment with PHF5A inhibitor, Pladienolide B reversed the induction of PHF5A overexpression and tumor formation of GC cells [9]. From this study, it is evident that AKT/mTOR, FOS and Skp2 interacts with PHF5A in GC and could be useful therapeutic targets.

**Table 1.** PHF5A involvement in cancers

| Cancer Type | Status | | | Cellular localization | Results | Expression levels of mRNA/protein after PHF5A depletion/loss | Reference |
|---|---|---|---|---|---|---|---|
| | In vitro | In vivo | Clinical samples | | | | |
| Lung adenocarcinoma (LAC) | LAC cell line (H1299 and H1975 cell lines) | Nude mice | Patients with LAC | Nucleus | PHF5A is overexpressed in LAC tissues and is significantly associated with tumor progression and poor patient prognosis. Knockdown of PHF5A resulted in reduced cell proliferation, cell arrest and suppressed migration and invasion of LAC cells | IGFBP3↑, DDIT3↑, P53↑ Skp2↓, AKT2↓, PI3K↓ | [12] |
| Lung adenocarcinoma (LAC) | H1299 and A549 cell lines | - | Patients with LAC | Nucleus | PHF5A promoted LUC progression by regulating the AS of cell cycle–associated genes (SKP2, CHEK2, ATR) and apoptosis–associated genes (API5 and BCL2L13) | SKP2↑, CHEK2↑, ATR↑ AP15↑, BCL2L13↑ | [13] |
| Non-small cell lung cancer (NSCLC) | NSCLC cell line (H1299 and A549 cell lines) | - | NSCLC tissue specimen | Nucleus | Knockdown of PHF5A diminished stemness phenotypes and HDAC8 expression | HDAC8↓, CD133↓, ALDH1↓ | [23] |
| Non-small cell lung cancer (NSCLC) | A549 cell lines | - | - | - | Knockdown of PHF5A downregulated ROCK and RhoA expressions | ROCK↓, RhoA↓ | [24] |
| Endometrial adenocarcinoma (EAC) | EAC derived cell lines and Non-malignant endometrium (NME) cell lines | Rat | Human EAC tumor samples | - | Changes in expression level for Phf5a/PHF5A in tumors of rat and human samples but observed pattern was inconsistent between the two species | - | [65] |
| Brain cancer | Glioblastoma multiforme stem cells (GSC) and Neural stem cells (NSC) cell lines | Nude mice | Patient-derived glioblastoma multiforme (GBM) stem cells (GSCs) | Nucleus | Knockdown of PHF5A disrupted splicing of multiple essential genes and induced cell cycle arrest and loss of viability in glioblastoma stem cell | PDGFRA↓, RAF1↓, HDAC6↓ | [7] |
| Colorectal cancer (CRC) | Human colon cancer cell lines (HCT8 and HCT116 cell lines) | Nude mice | CRC tissue specimen | Nucleus | PHF5A is overexpressed in CRC, associated with cell proliferation and metastasis and a poor patient prognosis. Knockdown of PHF5A induces genome-wide alternative splicing by inducing TEAD2 exon 2 inclusion to activate YAP signaling. Abnormal splicing of TEAD2 promotes proliferation and migration in CRC | TEAD2↓, BIRC5↓, ANKRD1↓, CYR61↓ | [26] |
| Colorectal cancer (CRC) | HEK293T and HCT116 cell lines | Nude mice | CRC tissue specimen | Nucleus | PHF5A hyperacetylation-induced alternative splicing to stabilize KDM3A mRNA and promotes KDM3A upregulation | - | [14] |
| Pancreatic cancer (PC) | Cancer stem cell (CSC) cell line (side population (SP) and non-side population (NSP) cell lines) | - | PC tissue specimen | Nucleus | PHF5A expression is upregulation in cell lines and PC tissues. Knockdown of PAF1 and PHF5A led to a decreased expression of NANOG | NANOG↓ | [11] |
| Hepatocellular carcinoma (HCC) | LO2 and HCC cell lines | | Human HCC tissues | Nucleus | PHF5A is upregulated in HCC tissues and cells. PHF5A downregulation inhibits the migration and invasion of HCC cells through NF-κB signaling pathways and inhibiting NF-κB signaling weakens PHF5A stimulatory effect on migration and invasion of HCC cells. | NF-κB ↓(p65↓, IKK-β↓, IκBα↓, MMP9↓, SLUG↓) | [10] |
| Gastric cancer (GC) | AGS and MGC-803 cell lines | Nude mice | Human GC tissues | Nucleus | PHF5A is overexpressed in GC and a poor prognosis indicator. PHF5A knockdown slows down proliferation, enhances apoptosis sensitivity and inhibit migration in GC cells. Knockdown of PHF5A in GC cells diminished PHF5A overexpression with decreased protein stability of FOS via Skp2-mediated ubiquitination. | p-AKT↓, p-mTOR↓<br>FOS↓, SKP2↑ | [9] |
| Breast cancer | Human breast epithelial MCF10A cells (Invasive breast carcinoma (CA1a) and carcinoma in situ (DCIS) cell lines) | NOD/ SCID mice | Breast cancer specimens | Nucleus | PHF5A is necessary for stabilizing SF3b spliceosome and complexes to histones. Knockdown of PHF5A decreased cell proliferation, migration, and tumor formation. Knockdown of PHF5A sensitized cancer cells to apoptotic signaling via AS-mediated FAS-activated serine/threonine kinase (FASTK) intron retention to increase apoptosis and decrease tumor progression in breast cancer. | FASTK-S↑ | [4] |

**Table 2.** Therapeutics targeting PHF5A

| PHF5A inhibition | Cancer type | Status | Status | Functional mechanism | Reference |
|---|---|---|---|---|---|
| | | **In vitro** | **In vivo** | | |
| Spliceostatin A, Sudemycin C1 or Pladienolide B | Brain cancer | + | - | PHF5A/U2 snRNP inhibition leading to splicing defects, cell arrest and loss of MYC mRNA expression | [7] |
| Pladienolide B | Colorectal cancer | + | - | Reversed the enhanced cell proliferation and migration caused by PHF5A overexpression. Inhibition of PHF5A-induced inclusion of TEAD2 exon 2 and the protein levels of TEAD2-L | [26] |
| Pladienolide B | Lung cancer | + | - | Inhibition of cell proliferation and induction of AS events in a dose–dependent manner similarly to PHF5A knockdown. | [13] |
| Pladienolide B | Gastric cancer | + | - | Reversed the induction of PHF5A overexpression on the malignant phenotypes and tumor formation of GC cells | [9] |

## 5. Future perspective

The role of PHF5A in regulating pluripotency, cell differentiation, and alternative splicing has been established, but there are still many aspects to explore to fully comprehend its significance in cancer development and progression. To shed light on the complexities of PHF5A and its potential as a therapeutic target and pave the way for innovative treatment approaches, the following future perspectives are suggested.

First, investigating the intricate molecular interactions of PHF5A with ATP-dependent helicases, the U2 snRNP spliceosome, and other targets such as CHD4, PAF1, and DDX will be crucial in unraveling their molecular interactions. Thus, understanding these interactions will provide insights into the specific mechanisms by which PHF5A influences splicing and its relationship with other pathways involved in cancer progression.

Secondly, further exploration of the genes and pathways affected by PHF5A-mediated alternative splicing could uncover potential therapeutic targets. By identifying specific splicing events regulated by PHF5A in different cancer types, it may be possible to develop interventions that restore normal splicing patterns or selectively target aberrant splicing events.

Furthermore, the tissue-specific expression of PHF5A suggests that it plays distinct roles in different organs. Therefore, investigating the tissue-specific functions of PHF5A in various cancers could provide valuable insights into organ-specific therapeutic strategies. Targeting PHF5A and its associated pathways within specific tissues may lead to more effective and tailored treatment approaches.

In addition, PHF5A has prognostic marker potential. Given the correlation between PHF5A expression and poor prognosis in several cancer types, further research could explore its utility as a prognostic marker. Understanding the prognostic value of PHF5A expression levels may help guide treatment decisions and personalize patient care.

Finally, developing PHF5A inhibitors or modulators could hold promise as targeted therapeutic options for cancer treatment. Building upon the knowledge gained from preclinical studies, investigating the efficacy and safety of potential PHF5A-targeting agents in clinical trials will be crucial to validate their therapeutic potential.

## 5. Conclusion

PHF5A is an oncoprotein involved in alternative splicing, histone deacetylase regulation, and signaling pathways. It promotes cancer growth and could be a potential therapeutic target.

**Credit Author Statement:** P.D.-N established review idea, information collection and writing. The author read and approved the final manuscript.

**Funding:** This research received no external funding.

**Informed Consent Statement:** Not applicable.

**Data Availability Statement:** Data used is included in the article.

**Acknowledgments:** Not applicable.

**Declaration of Interest:** The author declares no conflict of interest.